\RequirePackage{fix-cm}
\documentclass[smallextended]{svjour3}       
\smartqed  
\journalname{Mathematical Intelligencer}
\usepackage{graphicx,wrapfig,paralist}
\usepackage{url}
\def\cC{\mathcal{C}}
\def\cL{\mathcal{L}}

\author{Jacques Carette\and William M. Farmer\and Michael Kohlhase\and Florian
  Rabe}
\institute{Computing and Software, McMaster University\and
  Computing and Software, McMaster University\and
  Computer Science, FAU Erlangen-N\"urnberg\and
  Computer Science, FAU Erlangen-N\"urnberg}
\title{Big Math and the  One-Brain Barrier}
\subtitle{A Position Paper and Architecture Proposal}
\begin{document}
\maketitle
\begin{abstract}
  Over the last decades, a class of important mathematical results has required
  an ever increasing amount of human effort to carry out. For some, the help of
  computers is now indispensable.
  We analyze the implications of this trend towards \emph{big mathematics}, its
  relation to human cognition, and how machine support for big math can be organized.

  The central contribution of this position paper is an information model for \emph{doing mathematics}, which posits that humans very efficiently integrate five aspects of mathematics: \emph{inference}, \emph{computation}, \emph{tabulation},  \emph{narration}, and \emph{organization}.
  The challenge for mathematical software systems is to integrate these five aspects in the same way humans do.
  We include a brief survey of the state of the art from this perspective.

\smallskip
\noindent
\textbf{Acronyms}

\smallskip
\begin{tabular}{ll}  
CFSG:  & classification of finite simple groups\\
CIC:   & calculus of inductive constructions\\
DML:   & digital mathematics library\\
GDML:  & global digital mathematical library\\
LMFDB: & L-functions and Modular Forms Data Base\\
OBB:   & one-brain barrier\\
OEIS:  & Online Encyclopedia of Integer Sequences\\
OOT:   & Feit-Thompson Odd-Order Theorem\\
SGBB:  & small group brain-pool barrier
\end{tabular}
\end{abstract}


\section{Introduction}\label{sec:intro}

In the last half decade we have seen mathematics tackle problems that require increasingly
large digital developments: proofs, computations, data sets, and document collections. This trend
has led to intense discussions about the nature of mathematics, raising questions like
\begin{compactenum}[\em i\rm)]
\item Is a proof that can only be verified with the help of a computer still a
  mathematical proof?
\item Can a collection of mathematical documents that exceeds what can be understood in detail by a
  single expert be a legitimate justification of a certain mathematical result?
\item Can a collection of mathematics texts --- however big
  --- adequately represent a large body of mathematical
  knowledge?
\end{compactenum}
The first question was first raised by Appel and Haken's proof of the
four color theorem~\cite{AH:EPMIFC}, which in $400$ pages of regular
proof text reduced the problem to approximately $2000$ concrete configurations that had to be checked by a computer. Later it arose again from Thomas Hales's
proof of the Kepler conjecture~\cite{TH:APofKC}, which
contained substantial algebraic computations as part of the proof.
The second question is raised, e.g., for the classification of
finite simple groups (CFSG), which comprises the work of a large group
of mathematicians over decades and which has resisted dedicated
efforts to even write down consistently --- see the discussion below.
The third question comes from the ongoing development of digital
mathematics libraries (DMLs) --- such as the huge collection of papers
that constitute the CSFG --- that fail to make explicit the
abundant interconnections in mathematical knowledge that are needed to
find knowledge in these DMLs and reuse it in new contexts.

Let us call such developments \textbf{Big Math} by analogy to the
\emph{big data/big everything} meme but also alluding to the \emph{New Math}
movement of the 1960s that aimed to extend mathematical education by
changing how mathematics was taught in schools.  The emerging
consensus of the mathematical community seems to be that, while the
methods necessary for dealing with Big Math are rather problematic,
the results so obtained are too important to forgo by rejecting
such methods. Thus, we need to take on board Big Math methods and
understand the underlying mechanisms and problems.

In what follows, we analyze the problems, survey possible solutions, and propose a unified
high-level model that we claim computer support must take into account for scaling mathematics.
We believe that suitable and acceptable methods should be developed in a tight collaboration
between mathematicians and computer scientists --- indeed such method development is
already under way but needs to become more comprehensive and integrative.

We propose that all Big Math developments comprise five main aspects that need to be dealt with at scale:
\begin{compactenum}[\em i\rm)]
\item \emph{Inference}: deriving statements by \emph{deduction} (i.e.,
  proving), \emph{abduction} (i.e., conjecture formation from best
  explanations), and \emph{induction} (i.e., conjecture formation from examples).
\item \emph{Computation}: algorithmic manipulation and simplification
  of mathematical expressions and other representations of
  mathematical objects.
\item \emph{Tabulation}: generating, collecting, maintaining, and accessing
  sets of objects that serve as examples, suggest patterns and relations, and allow testing
  of conjectures. 
\item \emph{Narration}: bringing the results into a form that can be digested by humans in natural language but also in diagrams, tables, and simulations,
  usually in mathematical documents like papers, books, or webpages.
\item \emph{Organization}: representing, structuring, and interconnecting mathematical knowledge.
\end{compactenum} 
Computer support exists for all of these five aspects of Big Math, e.g., respectively,
\begin{compactenum}[\em i\rm)]
\item theorem provers like Isabelle~\cite{WLPN:TIF}, Coq~\cite{CoqManual}, or Mizar~\cite{mizar},
\item computer algebra systems like GAP~\cite{GAP}, SageMath~\cite{SageMath}, Maple~\cite{maple:url}, or Mathematica~\cite{mathematica:url},
\item mathematical data bases like the L-functions and Modular Forms Data Base (LMFDB)~\cite{Cremona:LMFDB16,lmfdb} or the Online Encyclopedia of Integer Sequences (OEIS)~\cite{Sloane:OEIS},
\item online journals, mathematical information systems like zbMATH~\cite{zbMATH:url} or MathSciNet~\cite{MathSciNet}, preprint servers like arXiv.org, or research-level help systems like MathOverflow~\cite{MathOverflow:web},
\item libraries of theorem provers and mathematics encyclopaedias like Wolfram MathWorld~\cite{URL:MathWorld} and PlanetMath~\cite{planetmath}.
\end{compactenum}
While humans can easily integrate these five aspects and do that for all
mathematical developments (large or otherwise), much research is still
necessary into how such an integration can be achieved in software systems. We
want to throw the spotlight on the integration problem to help start off
research and development of systems that integrate all five aspects.

Although the correctness of mathematical results is very important, checking
correctness is not our primary concern since there are effective societal
mechanisms and, when needed (as for Hales' proof of the Kepler conjecture),
inference systems for doing this.  Similarly, although the efficiency of
mathematical computation is likewise very important, computational efficiency
is also not our primary concern since there are programming languages and
computer algebra systems that enable highly efficient mathematical computation.
Correspondingly, there are highly scalable systems for working with large databases and collections of narrative documents.
Instead, our primary concern is to provide, by integrating these aspects, new
capabilities that are needed but currently unavailable for Big Math.

\paragraph*{Overview}\ 
In the next section we discuss some of the state of the art in computer support for Big Math by way 
 of high-profile mathematical developments, which we use as case studies for Big Math and present the 
 issues and methods involved.
In Section~\ref{sec:tetrapod}, we present a proposal for the integration of the five aspects into what we call a \emph{tetrapod}, whose body is  organization and whose legs are the other four aspects, arranged like the corners of a tetrahedron.
Section~\ref{sec:concl} concludes this position paper.

\section{Computer Support of Mathematics}\label{sec:computer-support}

The Classification of Finite Simple Groups (CFSG) is one of the seminal results
of 20\textsuperscript{th} century mathematics. Its usefulness and mathematical
consequences give it a prominent status in group theory, similar to that of the
fundamental theorem of arithmetic in number
theory. The proof of the CFSG was constructed through the coordinated effort of
a large community over a period of at least half a century; the last special
cases were only completed in 2004.

The proof itself is spread over many dozens of contributing articles summing up
to over 10,000 pages. As a consequence, work on collecting and simplifying the
proof has been under way since 1985, and it is estimated that the emerging
``second-generation proof'' can be condensed to 5000 pages~\cite{sg5000}.

It seems clear that the traditional method of doing mathematics, which
consists of well-trained, highly creative individuals deriving insights in the small, reporting on them in community meetings, and publishing them
in academic journals or monographs is reaching the natural limits posed by the
amount of mathematical knowledge that can be held in a single human brain ---
we call this the one-brain barrier\footnote{It might be argued that much
mathematical research is now carried out in small groups instead of by
individuals and that this should rather be called ``small group brain-pool
barrier'' (SGBB), but there are natural limits to collaboration on complex
topics, as has been epitomized in the seminal 1975 book ``\emph{The Mythical
Man-Month}''~\cite{Brooks:tmmm75}.  The main result of this study is that
adding members to a team can even slow down progress, because it induces
communication overhead; the main solution proposed is to introduce individuals
to the team that achieve ``a detailed understanding of the whole project'';
making the difference between an OBB and a SGBB gradual rather than
fundamental.} (OBB).

We posit that transcending the OBB will be a crucial step towards future
mathematics.  The space of mathematical knowledge on this side of the OBB is
bounded by the amount of time and memory capacity that a single individual
can devote to learning the scaffolding necessary to understand or build on a specific result.
More specifically, the point at which a mathematical domain grows so much that
it would take a mathematician more than $25$ years of work to
be able to contribute new ideas would likely signify the end of research in
that domain.  Indeed, we are seeing a gradual increase of large proofs, which
might point to a decrease of open important mathematical results inside the OBB. For example,
many high-profile open problems like the Riemann conjecture might be
elusive precisely because they are beyond the OBB.

There are two obvious ways around the OBB: 
\begin{compactenum}
\item breakthroughs in the structural
understanding of wide swaths of mathematics so that the effort of learning about
a particular domain can be greatly reduced; and 
\item large scale computer support.
\end{compactenum}
These are not mutually exclusive, and computer support may indeed enable such
breakthroughs.

\subsection{Computers vs. Humans in Mathematics}\label{sec:computers-humans}

Humans and computers have dual performance characteristics: Humans excel at
\textbf{vertical tasks} that involve deep and intricate manipulations,
intuitions, and insights but limited amounts of data. In contrast, computers
shine where large data volumes, high-speed processing, relentless precision,
but shallow inference are required: \textbf{horizontal tasks}.

In mathematics, vertical tasks include the exploration, conceptualization, and intuitive
understanding of mathematical theories and the production of mathematical
insights, conjectures, and proofs from existing mathematical knowledge.
Horizontal tasks include the verification of proofs, the processing of large
lists of examples, counterexamples, and evidence, and information retrieval
across large tracts of mathematical knowledge.

Enlisting computers for
horizontal tasks has been extremely successful --- mathematicians routinely use
computer algebra systems, sometimes to perform computations that have pushed
the boundary of our knowledge. Other examples include data-driven projects like
LMFDB, which tries to facilitate Langland's
Program~\cite{BerGel:ilp03} in number theory by collecting and curating objects
like elliptic curves and their properties, or OEIS, which similarly collects sequences of integers.
These already form Big Computation resp.
Big Tabulation (Big Data) approaches, but they do not help in the case of the CFSG,
which is more of a Big Inference and Big Narration problem though it involves
the other aspects as well.
In the sequel, we consider the issues
involved in an exemplary Big Inference effort.

\subsection{A Computer Proof of the Odd-Order Theorem}\label{sec:oorder}

In 2014, Georges Gonthier's team presented a machine-checked proof of the
Feit-Thompson Odd-Order Theorem (OOT) in the Coq theorem
prover~\cite{GonAspAiv:mcpoot13}. Even though Gonthier characterizes the OOT as
the ``\emph{foothills before the Himalayas constituted by the CFSG}'', the Coq
proof was a multi-person-decade endeavour, and the Coq verification ran multiple
hours on a current computer. This proof arguably sits at the edge of the OBB or maybe already transcends it.  In
this article, we want to analyze the kind of system we would need for
breaking the OBB and pushing the boundaries of mathematical knowledge. But
before we do, let us recap how proof verification works.

In a nutshell, the theorem and all the prerequisite background knowledge are
expressed in a logic, i.e., a formal language $\cL$ equipped with a calculus
$\cC$. In $\cL$ the well-formedness of an expression is rigorously
defined, so that it can be decided by running a program in finite time. $\cC$ is a set of rules that transform $\cL$-expressions
into other $\cL$-expressions, and a \emph{proof} of a \emph{theorem} $t$ (an
$\cL$-expression) is a series of applications of $\cC$-rules to $\cC$-axioms that
end in $t$. Essentially, a $\cC$-proof gives us
absolute assurance that $t$ is a theorem in $\cC$.
Crucially, the property of being a $\cC$-proof is also decidable.
However, the existence of a proof is usually undecidable, and the cost of producing a
proof is significant since its structure must be made explicit enough so that a
machine can fill all gaps using both decision procedures and heuristic search. Of course all lemmas
that lead up to the theorem are checked as well so that --- unlike in
informal mathematics --- we are sure that all pieces fit together exactly.

\subsection{The Engineering Aspect: Inference and Knowledge Organization}\label{sec:eng}

But this is only half the story.
To cope with the complexity of calculus-level proofs\footnote{Proofs in mathematics, on the other hand, are expressed in a natural language: mathematical vernacular~\cite{DeBruijn:tmv94}, a stylized form of English with interspersed mathematical formulae, tables, and diagrams.}, it is much more convenient to use expressive logics --- the calculus of inductive constructions (CIC)~\cite{BertotCasteran:coq04} in the case of Coq --- and programs that support the user in proof construction.
A proof like the one for the OOT can have billions of steps and is only ever generated in memory of the Coq system during proof checking.
Programs like Coq are engineering marvels, optimized to cope with such computational loads.

Optimization is also needed in the organization of the knowledge if one
is going to achieve the scale necessary for the OOT. Without care, we
frequently end up re-proving similar lemmas, resulting in an
exponential blow-up of the work required. To alleviate this problem, we follow
the (informal) mathematical practice of generalizing results and proving any
lemma at the most general level possible. However, in the formal methods
setting, we need to extend the logics, calculi, and proof construction
machinery involved to take modular development into account and optimize them
accordingly. For the OOT, Gonthier and his team developed the method of
\emph{mathematical components}~\cite{TasMah:mc:url} (akin to object-oriented modeling
in programming languages, but better suited to mathematics) inside CIC and
used that to control the combinatorics of mathematical knowledge
representation. Indeed, the development of the library of reusable
intermediate results comprised about 80\% of the development effort in the OOT
proof~\cite{GG:pc18}.

\section{Five Aspects of Big Math Systems}\label{sec:tetrapod}

We have seen two essential components of computer systems that can scale up to the Big Math level: 1.~efficient and expressive
theorem proving systems and 2.~systems for organizing mathematical knowledge in
a modular fashion.  Already in the introduction, we mentioned the five basic
\emph{aspects} of mathematics:
\begin{compactenum}[\em i\rm)]
\item \emph{Inference}, i.e., the acquisition of new knowledge from
 what is already known;
\item \emph{Computation}, i.e., the algorithmic transformation of
  representations of mathematical objects into more readily
  comprehensible forms;
\item \emph{Tabulation}, i.e., the creation of static, concrete data
  pertaining to mathematical objects and structures that can be
  readily stored, queried, and shared.
\item \emph{Narration}, i.e., the human-oriented description of
  mathematical developments in natural language; and
\item \emph{Organization}, i.e., the modular structuring of
  mathematical knowledge.
\end{compactenum}

\begin{figure}[ht]\centering
\includegraphics[width=5cm]{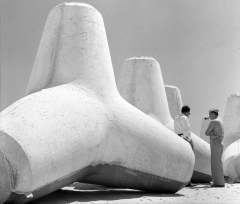}\qquad
\includegraphics[width=6cm]{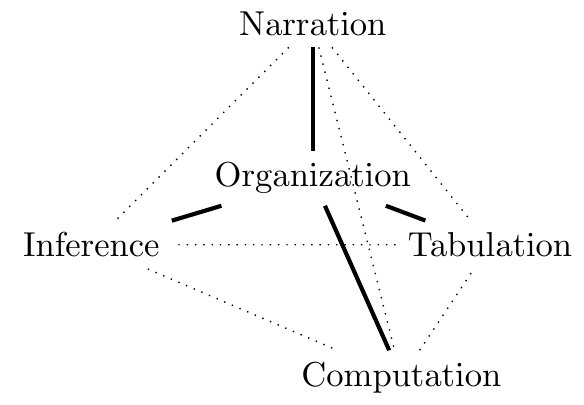}\qquad
\caption{The Five Aspects of Big Math Systems as a Tetrapod}\label{fig:tetrapod}
\end{figure}

These aspects --- their existence and importance to mathematics ---
should be rather uncontroversial. In order to help understand their tight relation, Figure~\ref{fig:tetrapod} arranges them in a convenient representation in three dimensions:
we locate the organization aspect at the centre and the other four aspects
at the corners of a tetrahedron since the latter are all consumers and producers of the mathematical knowledge represented by the former.
A four-dimensional representation might be more accurate but less
intuitive. We note that the names of the aspects are all derived from
verbs describing mathematical activity:
\begin{inparaenum}[\em i\rm)]
\item inferring mathematical knowledge, 
\item computing representations of mathematical objects,
\item tabulating mathematical objects and structures,
\item narrating how mathematical results are produced, and 
\item organizing mathematical knowledge.
\end{inparaenum}

Below we look at each aspect in turn using the CSFG and related efforts as
guiding case studies and survey existing solutions with respect to the tetrapod
structure from Figure~\ref{fig:tetrapod}.

\subsection{Inference}\label{sec:inference}

We have already seen an important form of machine-supported mathematical
inference: \emph{deduction} via machine-verified proof. There are other forms:
automated theorem provers can prove simple theorems by systematically searching
for calculus-level proofs (usually for variants of first-order logic), model
generators construct examples and counterexamples, and satisfiability solvers
check for Boolean satisfiability.
All of these can be used to systematically explore the space of mathematical knowledge and can thus constitute a horizontal complement to human facilities.

Other forms of inference yield plausible conclusions instead of provable facts:
\emph{abduction} (i.e., conjecture formation from best explanations) and \emph{induction} (i.e., conjecture formation from examples).
Machine-supported abduction and induction have been studied much less than machine-supported deduction, at least for producing formal mathematics. However, there is now a conference series~\cite{AITP:url} that studies the use of machine learning techniques in theorem proving.

One of the main problems with Big Inference for mathematics is that inference
systems are (naturally and legitimately) specialized to a
particular logic. For instance, interactive proof assistants like Coq and HOL
Light~\cite{Harrison:hlti96} have very expressive languages, whereas automated proof search is only
possible for simpler logics where the combinatorial explosion of the proof
space can be controlled. This makes inference systems very difficult to
inter-operate, and thus all their libraries are silos of formal mathematical
knowledge, thence leading to duplicated work and missed synergies --- in
analogy of the OBB we could conceptualize this as a \emph{one-system barrier of
formal systems}.  There is some work on logic- and library-level
interoperability --- we speak of \emph{logical pluralism} --- using meta-logical
frameworks (i.e., logics to represent logics and their
relations)~\cite{Pfenning:lf01,KohRab:qrtpflmk15,Saillard:tcmtp15}.
We contend that this is an important prerequisite for organizing mathematical knowledge in Big Math.

\subsection{Computation}\label{sec:computation}

Computer scientists have a very wide view of what \emph{computation} is:
be it $\beta$ reduction (in the case of the lambda calculus), transitions
and operations on a tape (for Turing machines), or rewrites in some symbolic language, all of these
are somehow quite removed from what a mathematician thinks when ``computing''.
Here, for the sake of simplicity and familiarity, we will largely be concerned
with \emph{symbolic computation}, i.e. manipulation of expressions containing
symbols that represent mathematical objects. Of course, there are also many 
flavours of numeric computation such as in scientific
computing, simulation/modelling, statistics, and machine learning.

In principle, mathematical computation can be performed by inference, e.g., by building a
constructive proof that the sum $2+2$ exists.
But this is not how humans do it --- they are wonderfully flexible in
switching between the computational and the inferential aspect.  Current inference-based systems in wide use have not achieved this flexibility,
although systems like Coq,
Agda~\cite{Norell:tppldtt07,AgdaWiki:url}, and Idris~\cite{idris:doc:url} are
making inroads.
In any case, computation via inference is intractable
(even basic arithmetic ends up in an unexpected complexity class) --- somewhat dual to how inference via computation in decision procedures has only limited success.

Instead, the most powerful computation systems are totally separate from inference: computer algebra
systems like Maple, Mathematica, SageMath, or GAP can tackle computations that are many
orders of magnitude larger than what humans can do --- often in mere milliseconds.

But these systems face the same interoperability problems as inference
systems do, open standards for object representation like
OpenMath \cite{OpenMath:url} or MathML~\cite{CarlisleEd:MathML3}
notwithstanding.  Just to name a trivial but symptomatic example, a particular
dihedral group is called $D_4$ in Sage and $D_8$ in GAP due to differing
conventions in the respective communities.
More mathematically involved and therefore more difficult to fix is that most of the implementations of special
functions in computer algebra systems differ in the underlying branch
cuts~\cite{CorDavJefWat:aas00}.  Inference during computation would enable some
of these problems to be fixed, but this has been sacrificed in favor of
computational efficiency. Another source of complexity is that today's most
feature-full symbolic computation systems are both closed-source and commercial,
which makes integrating them into a system of trustable tools challenging.
Having said that, the kinds of effort devoted to the development of those
systems is significantly higher than what can currently be achieved in academia, where code contributions are under-valued compared
to the publication of research papers. Furthermore, obtaining funding for
sustainable development of large software is difficult.

Lastly, there is the question of acceptability of certain computations in proofs.
In part, this derives from the difficulty of determining if a program written in a mainstream programming language is actually correct, at least to the same level of rigor that other parts of mathematics are subjected to.
Some of this problem can be alleviated by the use of more modern programming languages that have well understood operational and denotational semantics, thus putting them on a sound mathematical footing.
Nevertheless it will remain that any result that requires thousands of lines of code or hours of computation (or more) that cannot be verified by humans is likely to be doubted unless explicit steps are taken to insure its correctness.
The Flyspeck project~\cite{HalAdaBau:fpkc17} --- a computer verification of Thomas Hales' proof of the Kepler conjecture~\cite{TH:APofKC} in the HOL Light and Isabelle proof assistants of comparable magnitude to the OOT proof effort --- provides an interesting case study as it includes the verification of many complex computations.

\subsection{Tabulation}\label{sec:tabulation}

If we look at the CFSG, then already that result contains an instance of
tabulation: the collection of the 26 sporadic groups, which are concrete
mathematical objects that can be represented, e.g., in terms of matrices of numbers. Even more
importantly, many of the insights that led to the CFSG were only reached by
constructing particular groups, which were tabulated (as parts of
journal articles and lists that were passed around).
We also see this in other Big Math projects, e.g., Langland's program of trying to relate Galois groups
in algebraic number theory to automorphic forms and the representation theory of
algebraic groups over local fields and adeles. This is supported by LMFDB,
which contains about 80 tables with representations of mathematical objects
ranging from elliptic curves to Hecke fields --- almost a terabyte of data all
in all. These are used, e.g., to find patterns in the objects and their
properties and to support or refute conjectures. Other well-known examples are the OEIS with over 300,000
integer sequences and the Small Groups Library~\cite{SmallGroups:url} with more than 450 Million groups of small order.
See \cite{Bercic:cmo:wiki:url} for a work in progress survey of math databases.

Unfortunately, such math databases are typically not integrated with systems for
mathematical inference, narration, or knowledge organization and
only weakly integrated with systems for computation. Usually these
databases supply a human-oriented web interface.
If they also offer API access to the underlying database, they only do so for database-level interaction, where an elliptic curve might be a quadruple of numbers encoded as decimal strings to work around size restrictions of the underlying
database engine.
What we would need instead for an integration in a Big Math system would be an API that supplies access to the mathematical objects ---
e.g., the representations of elliptic curves as they are treated in organization, computation, inference, and narration systems, see~\cite{WieKohRab:vtuimkb17} for a discussion.

As usual, there are exceptions.
The GAP Small Groups Library system and LMFDB are notable examples: the former is deeply integrated with the GAP computer algebra system, and the latter includes almost a thousand narrative descriptions of the mathematical concepts underlying the tabulated objects.

\subsection{Narration}\label{sec:narration}

\begin{figure}[ht]\centering
  \includegraphics[width=.98\textwidth]{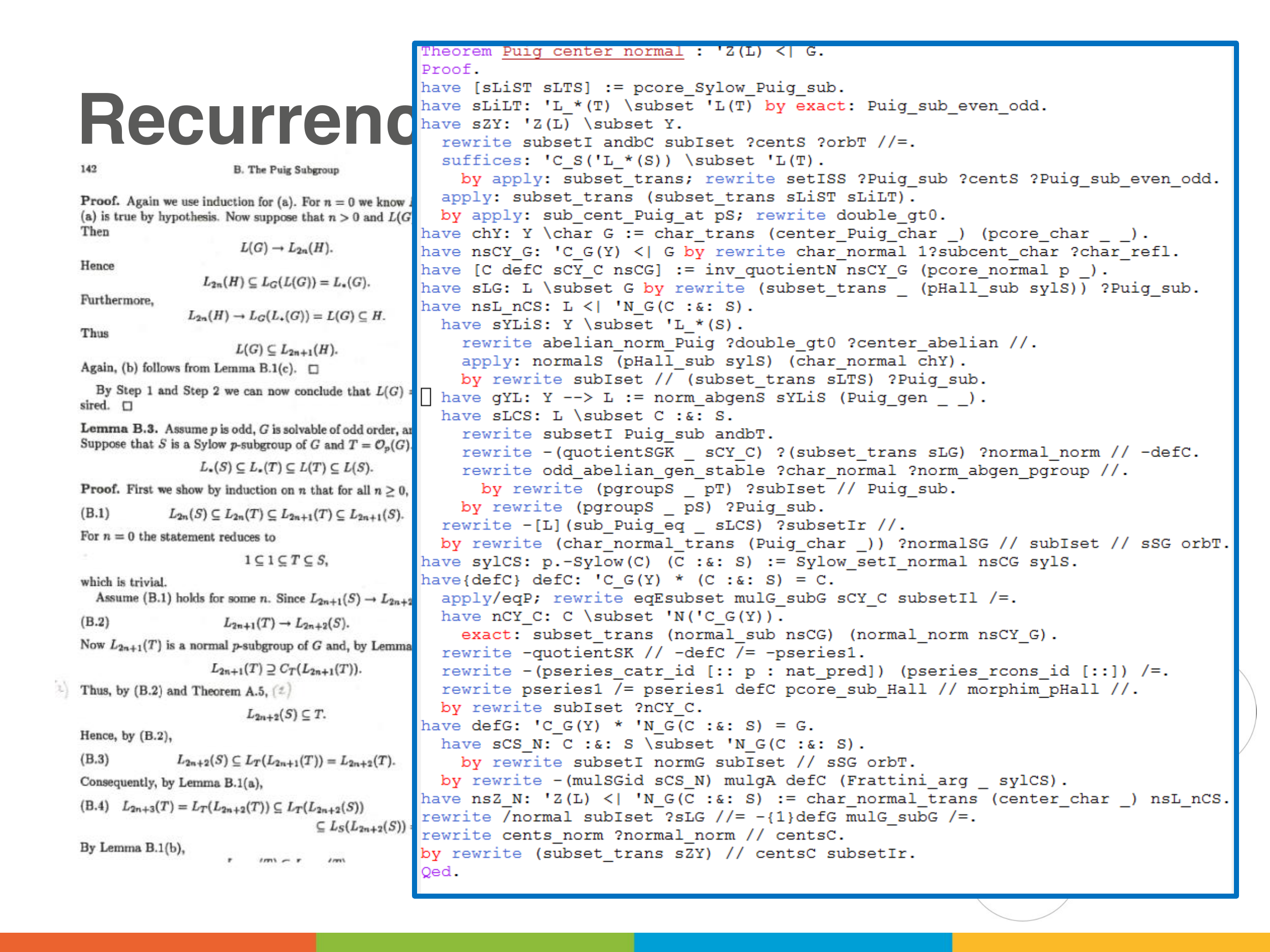}
  \caption{Informal and Formal Representations of Mathematical Knowledge}\label{fig:puig}
\end{figure}

Consider Figure~\ref{fig:puig}, which shows an intermediate result in the OOT in the Coq inference system (foreground) and its corresponding narrative representation (background).
Even though great care has been taken to make the Coq text human-readable, i.e., short and suggestive of traditional mathematical notation, there is still a significant language barrier for all but the members of the OOT development team.

Indeed, mathematical tradition is completely different from its representation in inference, computation, and tabulation systems.
Knowledge and proofs are presented in documents like journal articles, preprints, monographs, and talks for human consumption.
While rigour and correctness are important concerns, the main emphasis is on the efficient communication of ideas, insights, intuitions, and inherent connections to colleagues and students.
As a consequence, more than half of the text of a typical mathematical document consists of introductions, motivations, recaps, remarks, outlooks, conclusions, and references.
Even though this packaging of mathematical knowledge into documents leads to extensive overhead and duplication, it seems to be an efficient way of dealing with the OBB and thus a necessary cost for scholarly communication.

In current proof assistants like Coq, the narration aspect is under-supported even though tools like {\LaTeX} have revolutionized mathematical writing.
Source comments in the input files are possible in virtually all inference and computation systems, but they are
not primary objects for the system and are thus used much less than in narrative representations.
Knuth's \emph{literate programming} idea~\cite{DK:LP} has yet to take root in mathematics although it is worthwhile noting that one of the earliest inference systems, Automath~\cite{DeBruijn-70-a}, had extensive features for narration.
The main modern exception among inference systems is Isabelle: it supports the inclusion of marked up text and programs and turns the underlying ML into a deeply integrated document management system, which allows recursive nesting of narrative, inference, and computation~\cite{Wenzel:IIdsflitd18}.

Some computational communities such as parts of statistics, especially users of R, make use of some features of literate programming.
Jupyter notebooks as used with SageMath as well as the document interfaces of Maple and Mathematica are also somewhat literate although the simple fact that they do not interoperate smoothly with {\LaTeX} hampers their adoption as methods of conveying large amounts of knowledge narratively.

In any case, inference and computation systems are notoriously bad for expressing the vague ideas and underspecified concepts that are characteristic for early phases of the development of mathematical theories and proofs, a task at which narration excels.
Therefore, the Flyspeck project~\cite{HalAdaBau:fpkc17} used a {\LaTeX}-based book~\cite{Hales:DenseSpherePackings12} that refactored the original proof to orchestrate and track the formal proof via cross-references to the identifiers of all formal definitions, lemmas, and theorems.
Incidentally, the ongoing effort of establishing a second-generation proof of the CSFG has a similar book, consisting of seven volumes already published and five additional volumes that will be published in the future~\cite{sg5000}.

\subsection{Organization}\label{sec:organization}

In the discussion and survey of the four corners of the tetrapod from
Figure~\ref{fig:tetrapod}, we have seen that all four aspects share and are based on
the representation of knowledge, which we call the \emph{mathematical ontology}\footnote{Note that we
will use the word ``ontology'' in the original wider of ``a set of concepts and categories in a subject area or domain that
shows their properties and the relations between them.'', not just for
specific technologies of the Semantic Web~\cite{URL:W3}.}, and that they can
interoperate effectively through this ontology.
And we have seen that a modular, redundancy-minimizing organization of the ontology is crucial for
getting a handle on the inherent complexity of mathematical knowledge.

Most inference and computation systems feature some kind of modularity features
to organize their libraries. For inference systems, this was pioneered in the IMPS
system~\cite{FaGu93} in the form of theories and theory interpretations and
has been continued, e.g., in the Isabelle system (type classes and locales). In systems like Coq
or Lean~\cite{MouKonAvi:tltp15} that feature dependent record types, theories
and their morphisms can be encoded inside the base logic.
Computation systems feature similar concepts.
Finally, the MMT system~\cite{RabKoh:WSMSML13,uniformal:url} systematically combines modular
representation with a meta-logical framework, in which the logics and
logic-morphisms can be represented themselves, yielding a
foundation-independent (bring-your-own-logic) framework for mathematical
knowledge representation that can be used to establish system interoperability.

\section{Conclusion}\label{sec:concl}

Using the classification of finite simple groups as an example of Big Math, we
diagnosed the one-brain barrier as a major impediment towards
large-scale developments and results in mathematics.  We saw that the seemingly obvious
answer to this problem --- employ computer support --- is not without problems and can be
a barrier itself. We proposed that computer-based mathematical assistants
should have a tetrapodal structure, integrating inference, computation,
tabulation, and narration centered around a shared knowledge organization feature. We claim that
only with special consideration of all these five aspects will mathematical software
systems be able to render the support that is necessary for Big Math
projects like the CFSG to go beyond the proof certification service
rendered by big formal proofs like OOT or Flyspeck.

While the holistic conception of Big Math envisioned by our tetrapod is new, the general
sentiment that mathematical assistant systems must escape their native corner
is implicitly understood in the mathematical software community. 
Indeed, many of the systems we mentioned above, while focusing on a particular
aspect and excelling at it, also integrate features of some other
aspects. We have briefly surveyed current efforts towards such integrations.
A thorough review of the state of the art, which would more clearly delineate the
progress on the roadmap implicitly given by the tetrapod proposal, is beyond the
scope of this position paper but is under active development by the authors
for future publication.

We have observed the central place of the ontology in the proposed system functionality
architecture, and we claim that such systems are best served by a global digital
mathematical library (GDML)~\cite{NAS14}, which serves as a pivotal point for integrating systems and
system functionalities.
Again, a discussion of this important resource is beyond the scope
of this paper, but see~\cite{Farmer:mifr04} for a motivation and~\cite{KohRab:qrtpflmk15}
for an avenue on how it could be realized by marshalling exiting resources like the
libraries of proof assistants.
\medskip

A GDML would constitute a critical resource for mathematics; it should thus be
unsurprising to find organization at the centre of the tetrapod.
Arguably, the comprehensive ontology of mathematics that would constitute Big Organization cannot be created by a small set of individuals --- it
is both subject to and a way around the one-brain-barrier --- but has to be a
collaborative effort of the whole community.  A prerequisite for this is that
the ontology be FAIR (Findable,
Accessible, Interoperable, and Reusable;~\cite{WilDumAal:FAIR16}), open, and
not encumbered by commercial or personal interests that are
insurmountable for researchers in developing countries.
This remains true even though revenue streams to fund the maintenance of such an ontology are necessary; thus, suitable licensing schemes and business models that reconcile the openness and funding considerations will have to be found.

Another trade-off to consider is that current commercial software development for mathematics
has been most effective in regards to large-scale integration projects ---
the various offerings by Wolfram Inc. (Mathematica and its
Notebooks, Wolfram Alpha, and the Wolfram Language) together constitute one of the most
tetrapodal system that currently exists.  Non-commercial funding for a
similar open effort simply does not seem to exist at the moment.  The downside of commercial
software and accompanying resources is that it often cannot be adapted for
research purposes without considerable legal and monetary effort.
No matter how useful it is, closed source projects ought to have the same impact as omitted proofs: greatly inspiring doubt.

Nevertheless, the authors believe that a confederated, well-organized community
effort to develop open-source software and open resources can succeed.  This
is well-supported by the current practices of the mathematical
community, who by and large embrace open communication, open-source
software, and open access publication of results and resources. 
This is not anti-commercialization as there are several very successful companies
thriving on open source software; and even closed-source behemoths like Microsoft and Google have increasingly released
part of their source code to the community.
\medskip

We acknowledge the fact that our tetrapod proposal does not incorporate the
fact that mathematics is a social process and
that for Big Math problems we will need mathematical social machines, i.e., ``an environment comprising humans and technology interacting and
  producing outputs or action which would not be possible without both parties present''~\cite{Berners-Lee:wtw99}.
We conjecture that the social machine aspect is one that will live quite
comfortably on top of tetrapod-shaped mathematical software systems and can
indeed not fully function without all of its five aspects interacting well;
see~\cite{Martin:cls16,CorMarMur:mwmad17} for a discussion and further pointers
to the literature.

Finally, we remark that of course the OBB is not particular to mathematics but
affects all scientific and engineering disciplines, where we conjecture
similar tetrapodal paradigms as ours apply.
Here, mathematics is a very good test case for the design of knowledge-based
systems since the knowledge structures and algorithms are so overt.

\paragraph*{Acknowledgments}\
The authors gratefully acknowledge fruitful discussions with many of our
colleagues in the mechanized mathematics community.  The work reported here was
supported by the Horizon 2020 European Research Infrastructures
project OpenDreamKit (\#676541), the DFG project RA-18723-1 OAF, and the NSERC Grants RGPIN-2018-05812
and RGPIN-5100-2015.

\newcommand{\etalchar}[1]{$^{#1}$}

\end{document}